\documentclass[aps,showpacs]{revtex4}
\usepackage{amsmath}
\usepackage{amssymb}
\usepackage{epsfig}
\usepackage[dvips]{color}

\begin{document}
\title{Location- and observation time-dependent quantum-tunneling}
\author{ V. Fleurov$^1$, R. Schilling$^2$, and B. Bayani$^2$}
\affiliation{$^1$Raymond and Beverly Sackler Faculty of Exact
Sciences, School of Physics and Astronomy,\\ Tel-Aviv University,
Tel-Aviv 69978 Israel.}
\affiliation{$^2$ Johannes Gutenberg University, Mainz, Germany}
\begin{abstract}
We investigate quantum tunneling in a translation invariant chain of
particles. The particles interact harmonically with their nearest
neighbors, except for one bond, which is anharmonic. It is described
by a symmetric double well potential. In the first step, we show how
the anharmonic coordinate can be separated from the normal modes.
This yields a Lagrangian which has been used to study quantum
dissipation. Elimination of the normal modes leads to a nonlocal
action of Caldeira-Leggett type. If the anharmonic bond defect is in
the bulk, one arrives at Ohmic damping, i.e. there is a transition
of a delocalized bond state to a localized one if the elastic
constant exceeds a critical value $C_{crit}$. The latter depends on
the masses of the bond defect. Superohmic damping occurs if the bond
defect is in the site $M$ at a finite distance from one of the chain
ends. If the observation time $T$ is smaller than a characteristic
time $\tau_M\sim M$, depending on the location $M$ of the defect,
the behavior is similar to the bulk situation. However, for $T \gg
\tau_M $ tunneling is never suppressed.
\end{abstract}
 \maketitle

\section{INTRODUCTION}

The influence of environmental degrees of freedom (DOF) on quantum
phenomena, like e.g. tunneling, has been of great interest during
the last decades \cite{1,2,3}. A significant progress came from the
investigation of a more or less phenomenological model where a
particle in a one-dimensional potential $V(q)$ or a free particle is
coupled to a bath of harmonic oscillators with the spectral density
$J(\omega)$. The quantum dissipation generated by the bath depends
qualitatively on the low frequency behavior of $J(\omega)$
\cite{1,2,3}. A particular interesting case is Ohmic damping, when
$J(\omega) \sim\omega$ for low enough frequencies. In that case and
for a symmetric double well potential, the particle at zero
temperature undergoes a transition from a delocalized state to a
localized one if the coupling constant between the particle and bath
exceeds a critical value \cite{4}. An interesting observation has
been made by Caldeira and Leggett \cite{5}. The exponent of the
exponential factor for the tunneling probability is multiplied by
$\eta$, the phenomenological friction coefficient of the
corresponding \textit{classical} dynamics. This relationship between
classical and quantum dissipation has been deepened and generalized
by Leggett \cite{6} for an arbitrary linear coupling between the
particle and bath coordinates.

Let $\tilde{q}(\omega)$ be the Fourier transform of the classical
particle trajectory $q(t)$ and
\begin{equation}
\widetilde{K}_0(\omega)\tilde{q}(\omega) + \frac{\widetilde{\partial
V}}{\partial q}(\omega) = 0\label{eq1}
\end{equation}
the transformed classical equation of motion, where
$\widetilde{K}_0(\omega)$ contains the dissipative influence of the
bath. Then the reduced Euclidean particle propagator $G_E (q',
T|q,0)$ (where the harmonic DOF have been eliminated) can be
represented by a path integral in the imaginary time $t = -i\tau$
\cite{7}
\begin{equation}
G_E(q',T|q,0) = \int\limits_{q(0)=q \atop
q(T)=q'}\mathcal{D}[q(\tau)]e^{-\frac{1}{\hbar} S[q(\tau)]}
\label{eq2}\qquad.
\end{equation}
The action
\begin{equation}
S[q(\tau)] = S_0[q(\tau)] +
S_{nonlocal}[q(\tau)]\tag{3a}\label{eq3a}
\end{equation}
contains the local
\begin{equation}
S_0[q(\tau)] = \int\limits_0^Td\tau\,
\left[\frac{M_p}{2}\dot{q}(\tau)^2+V(q(\tau)) +
\frac{\mu}{2}q(\tau)^2\right] \tag{3b} \label{eq3b}
\end{equation}
and the nonlocal part
\begin{equation}
S_{nonlocal}[q(\tau)] = - \int\limits_0^T
\,d\tau\int\limits_0^{\tau}\,d\tau'\, K(\tau-\tau')q(\tau)q(\tau') =
-\frac{1}{2\pi}\frac{1}{2}\int d\omega \,\widetilde{K}(\omega)|
\tilde{q}(\omega)|^2. \tag{3c}\label{eq3c}
\end{equation}
The second equality holds for $T\rightarrow \infty$.
$\widetilde{K}(\omega)$ is the Fourier transform of the integral
kernel $K(\tau)$ and it is related to $\widetilde{K}_0(\omega)$ by
$\widetilde{K}(\omega) = \frac{1}{2}\widetilde{K}_0(-i|\omega|)$. If
the Euclidean Lagrangian of the particle-bath system is
\begin{equation}
L = L_0 + L_1 \quad,\qquad L_1 = L_{bath} +
L_{int}\tag{4a}\label{eq4a}
\end{equation}
with
\begin{equation}
L_0(q,\dot{q}) = \frac{1}{2} M_p \dot{q}^2 +
V(q),\tag{4b}\label{eq4b}
\end{equation}
and
\begin{equation}
L_1(q,\underline{x};\dot{q},\underline{\dot{x}}) = \frac{1}{2}
\sum_{\alpha=1}^N m_{\alpha} \left[\dot{x}_{\alpha}^2 +
\omega_{\alpha}^2 \left(x_{\alpha} - \frac{c_{\alpha}}{m_{\alpha}
\omega_{\alpha}^2}q \right)^2 \right]\tag{4c} \label{eq4c}
\end{equation}
then\cite{1,2,3}
\begin{equation}
\mu = \frac{2}{\pi} \int\limits_0^{\infty} d\omega\,
\frac{J(\omega)}{\omega} \tag{5a}\label{eq5a}
\end{equation}
and
\begin{equation}
K(\tau) = \frac{1}{\pi} \int\limits_0^{\infty} d\omega\, J(\omega)
\frac{\cosh(\omega(\frac{T}{2} - |\tau|))}{\sinh(\omega
\frac{T}{2})}\tag{5b}\label{eq5b}
\end{equation}
with the spectral density
\begin{equation}
J(\omega) = \frac{\pi}{2} \sum_{\alpha=1}^N
\frac{c_{\alpha}^2}{m_{\alpha} \omega_{\alpha}} \delta(\omega -
\omega_{\alpha})\qquad .\tag{5c}\label{eq5c}
\end{equation}
For a finite $T$ the kernel $K(\tau)$ and the paths $q(\tau)$ can be
periodically continued. Then the Fourier series for $K(\tau)$ is
given by the Fourier coefficients \cite{3}
\setcounter{equation}{5}
\begin{equation}
K_n = \frac{1}{T} \sum_{\alpha} \frac{c_{\alpha}^2}{m_{\alpha}}
\frac{1}{\nu_n^2+\omega_{\alpha}^2}\label{eq"1"}
\end{equation}
with $\nu_n = (2\pi /T)n,\quad n=0, \pm 1, \pm 2, ...$. Here $M_p$
and $\{m_\alpha\}$ are the masses of the particle and harmonic
oscillators, respectively. $\{\omega_\alpha\}$ are the oscillator
frequencies and $\{c_\alpha\}$ are the coupling constants between q
and the coordinates of the oscillators $\{x_\alpha\}$. Note that
$\{x_\alpha\}$ are not necessarily positions, but can represent
normal mode coordinates of vibrations, etc.

In the following we will restrict ourselves to a system of $N$
particles whose potential energy $V(\vec{x}_1,\ldots ,\vec{x}_N)$
includes harmonic and anharmonic interactions. Without an external
field, $V$ must be translationally invariant. However, since the
coordinates of Langrangian (4) are not specified, it is not
necessarily invariant under translations. This has motivated
Chudnovsky \cite{8} to apply the Caldeira-Leggett approach to a
system of two particles ($i = 1,2$) with positions $x_i$ and masses
$M_i$, coupled to oscillators with coordinates $x_\alpha$,
frequencies $\omega_{\alpha}$ and masses $m_\alpha,\alpha=1,\ldots
,N$. The corresponding Euclidean Lagrangian is of the form
 \begin{equation}
L = \frac{M_1}{2}\dot{x}_1^2 + \frac{M_2}{2}\dot{x}_2^2 + V(x_1-x_2)
+ \frac{1}{2} \sum_{\alpha=1}^N m_{\alpha} \left[\dot{x}_{\alpha}^2
+ \omega_{\alpha}^2(x_{\alpha} - x_2)^2\right] \label{eq6}
\end{equation}
with $V(x_1-x_2)$ being the interaction energy between both
particles. The coupling of particle $i = 1$ to the oscillators is
assumed to be zero. It is obvious that $L$ is translationally
invariant under $x_i \longrightarrow x_i + a\,,\quad
x_{\alpha}\longrightarrow x_{\alpha} + a$ provided that
$\{x_{\alpha}\}$ are real space coordinates. Surprisingly the
elimination of harmonic DOF does not lead for zero temperature to a
nonlocal action of Caldeira-Leggett type (cf. Eqs. (\ref{eq3c}),
(\ref{eq5b}),(\ref{eq"1"})). The Fourier coefficients of
Chudnovsky's kernel have the form
\begin{equation}
K_n^c = \frac{2}{T} \frac{M_1^2\nu_n^2}{M_1+M_2 + \displaystyle
\sum_{\alpha} \frac{m_{\alpha}\omega_{\alpha}^2}{\nu_n^2 +
\omega_{\alpha}^2}},\label{eq"2"}
\end{equation}
which differs qualitatively from the Caldeira-Leggett type result
(\ref{eq"1"}). However, for $M_1\rightarrow\infty$ one gets
\begin{equation}
K_n^c\rightarrow\frac{2}{T} (M_1-M_2)\nu_n^2-\frac{1}{T}
\sum_{\alpha}2m_{\alpha}\omega_{\alpha}^2
\frac{\nu_n^2}{\nu_n^2+\omega_{\alpha}^2}\label{eq"3"}
\end{equation}
where the last term, up to the additional factor $-\nu_n^2$,
corresponds to the second order time derivative of the kernel,
identical to $K_n$ from Eq. (\ref{eq"1"}), if one chooses
$c_{\alpha}^c = \sqrt{2}m_{\alpha}\omega_{\alpha}$\cite{8}.

Then the question arises: Does a translationally invariant model in
general lead to a nonlocal action, which is not of Caldeira-Leggett
type? This is one of the main points we want to investigate among
others in our paper. It will be done for a microscopic(within a
Born-Oppenheimer approximation), explicitly translationally
invariant lattice model with a defect which cannot diffuse. We will
show how the normal mode coordinates for the harmonic DOF can
exactly be separated from the anharmonic ones. This leads to a
Lagrangian of the form of Eq.~(4) with coupling constants
$c_\alpha$, frequencies $\omega_\alpha$ and a spectral density
determined by the microscopic model parameters. Such a microscopic
justification of Lagrangian (4) has been presented for quantum
\textit{diffusion} \cite{9}. There a particle diffusing through an
elastic lattice is considered. If, however, that particle cannot
diffuse and is an integral part of the lattice, e.g. an impurity
which can tunnel only between two positions, one has to separate the
center of mass (COM) and relative coordinates of \textit{all}
particles. For such a situation, Sethna \cite{10} has estimated the
coupling constants $c_\alpha$ by comparing the strain field of an
elastic monopole of an impurity with the displacement for a
longitudinal mode. But, as far as we know, there is no microscopic
derivation available for the quantities $c_\alpha$, $\omega_\alpha$
and $J(\omega)$ for a non-diffusing impurity in a lattice. Besides
such a microscopic derivation we will show that the quantum behavior
of the defect is sensitive to its location.

The outline of our paper is as follows: Section \ref{Model} presents
our model and outlines the main steps leading to Lagrangian (4). The
implications for the quantum behavior will be discussed in the third
section with a special emphasis on the role of defect location. A
summary and conclusions are contained in the final section
\ref{summary}. Appendices A, B and C contain details on the
separation of the harmonic and anharmonic DOF.

\section{Model}
\label{Model}

We consider an open chain of $N$ particles with masses $m_n,\ n=1,
\ldots ,N$ and harmonic nearest neighbor interactions and one
anharmonic bond representing a defect. This model could describe a
linear macromolecule with an impurity. The classical Hamiltonian
reads
\begin{equation}
H = \sum_{n=1}^N\frac{1}{2m_n}p_n^2 + V(x_1, ...,
x_N)\tag{10a}\label{eq10a}
\end{equation}
with the potential energy
\begin{equation}
V(x_1, ..., x_N) = \frac{C}{2}\sum_{n=1 \atop (\neq
M)}^{N-1}(x_{n+1}-x_n-a)^2 + V_0(x_{M+1}-x_{M})\qquad
.\tag{10b}\label{eq10b}
\end{equation}
in which $x_n$ is the position of $n$-th particle, $C$ is the
elastic constant of the harmonic nearest neighbor interaction, $a$
is the equilibrium length of the harmonic bonds and $V_0 (x_{M+1} -
x_M)$ is the energy of the anharmonic bond defect located between
the sites $M$ and $M + 1$. The potential energy $V$ is explicitly
translationnally invariant. Several ways exist to separate the
harmonic and anharmonic DOF, which finally lead to the Langrangian
(4) (see Appendices \ref{B} and \ref{C}). One (see Appendix
\ref{B}), which is also applicable in higher dimensions, is to
introduce the COM
\begin{equation}
X_d = \frac{1}{m_M + m_{M+1}}\left(m_Mx_M + m_{M+1} x_{M+1}\right)
\tag{11a}\label{eq11a}
\end{equation}
and relative coordinates of the \textit{bond defect}
\begin{equation}
q_M = x_{M+1}-x_M\qquad .\tag{11b}\label{eq11b}
\end{equation}
Then, $x_1,\ldots ,x_{M-1},X_d,x_{M+1},\ldots ,x_N$ are harmonic
coordinates. Their kinetic and potential energy can be diagonalized
by introducing normal coordinates. $q_M$ is linearly coupled to
these coordinates. As a result, one obtains the Lagrangian (4). Here
we will choose a different approach applicable to 1d systems, which,
however, leads to the same Lagrangian. Let
\begin{equation}
X_c = \frac{1}{M_c}\sum_{n=1}^Nm_n x_n\,,\quad M_c =
\sum_{n=1}^{N}m_n\tag{12a}\label{eq12a}
\end{equation}
be the COM of \textit{all} the particles and
\begin{equation*}
q_i = x_{i+1} - x_i - a_i\,,\quad i=1, ..., N-1,
\end{equation*}
\begin{equation}
a_M = 0\,,\quad \mbox{and}\ a_i = a\, \quad \mbox{otherwise,}
\tag{12b}\label{eq12b}
\end{equation}
be the relative coordinates, respectively. Using notation $q_0 = X_c$,
Eq.~(\ref{eq12a}),(\ref{eq12b}) has the form
\begin{equation}\label{eq13a}
q_i + a_i= \sum_{n=1}^{N} A_{in}x_{n}\,,\quad i = 0,1, ...,
N-1\qquad .\tag{13a}
\end{equation}
with
$$
A_{in} = \frac{m_{n}}{M_c} \delta_{0,i} + (\delta_{i,n+1} -
\delta_{i, n})(1-\delta_{0,i})
$$
Let $\pi_i$ be the canonical conjugate momenta of $q_i$. It is easy
to prove that Eq.~(\ref{eq13a}) implies
\begin{equation}
p_n = \sum_{i = 0}^{N-1} A_{in}\pi_{i}\,,\quad n = 1, ..., N\qquad
.\tag{13b}\label{eq13b}
\end{equation}
Substituting $p_n$ into Eq.~(\ref{eq13a}) yields
\setcounter{equation}{13}
\begin{eqnarray}
H & = & \frac{1}{2M_c}\pi_0^2 + \frac{1}{2} \sum_{i=1}^{N-1}
\left(\frac{1}{m_i} + \frac{1}{m_{i+1}}\right)\pi_i^2 -
\sum_{i=1}^{N-2}\frac{1}{m_{i+1}}\pi_i\pi_{i+1}\nonumber
\\
&\phantom{+}& + \frac{C}{2}\sum_{i=1 \atop (i\neq M)}^{N-1}q_i^2 +
V_0(q_M)\qquad .\label{eq11}
\end{eqnarray}
Here the first term is the kinetic energy of COM, which will be
dropped from now on. Note that the use of relative coordinates
introduces a coupling between the momenta. In the next step, we
perform the canonical transformation
\begin{equation*}
\pi_i = \tilde{p}_i\,,\quad i = 1, ..., N-1;\, i\neq M,
\end{equation*}
\begin{equation*}
 \pi_M = \tilde{p}_M + \frac{1}{m_M +
m_{M+1}}\left(m_{M+1}\tilde{p}_{M-1} +
m_M\tilde{p}_{M+1}\right),\tag{15a}\label{eq15a}
\end{equation*}
\begin{equation*}
q_i = \tilde{q}_i\,,\quad i=1, ..., N-1;\, i\neq M\pm
1
\end{equation*}
\begin{equation}
q_{M-1} = \tilde{q}_{M-1} - \frac{m_{M+1}}{m_M + m_{M+1}}
\tilde{q}_M,\tag{15b} \label{eq15b}
\end{equation}
\begin{equation*}
q_{M+1} = \tilde{q}_{M+1} - \frac{m_M}{m_M +
m_{M+1}}\tilde{q}_M\qquad .
\end{equation*}
This leads to
\begin{equation}
H = H_d + H_{harm} + H_{int}\tag{16a}\label{eq16a}
\end{equation}
where
\begin{equation}
H_d = \frac{m_M + m_{M+1}}{2m_Mm_{M+1}}\tilde{p}_M^2 +
V_0(\tilde{q}_M) + \frac{C}{2}\frac{m_M^2 + m_{M+1}^2}{(m_M +
m_{M+1})^2}\tilde{q}_M^2\tag{16b}\label{eq16b}
\end{equation}
is the defect Hamiltonian,
\begin{equation}
H_{harm} = \frac{1}{2}\sum_{i,j=1\atop (\neq M)}^{N-1}
T_{ij}\tilde{p}_i\tilde{p}_{j} + \frac{C}{2}\sum_{i=1\atop (\neq
M)}^{N-1} \tilde{q}_i^2\qquad .\tag{16c}\label{eq16c}
\end{equation}
is the harmonic part of the Hamiltonian and
\begin{equation}
H_{int} = -C\frac{m_{M+1}\tilde{q}_{M-1} + m_M\tilde{q}_{M+1}}{m_M +
m_{M+1}} \tilde{q}_M\tag{16d}\label{eq16d}
\end{equation}
is the coupling between the two.

The matrix $\textbf{T}=(T_{ij})$ in (\ref{eq16c}) depends on the
masses $m_i$ (see Appendix \ref{A}). Let $\textbf{e}^{(\alpha)} = (
e_1^{(\alpha)}, \ldots , e_{M-1}^{(\alpha)}, e_{M+1}^{(\alpha)},
\ldots , e^{(\alpha)}_{N-1})^t$ and $\lambda _ \alpha$ be the
normalized eigenvectors and eigenvalues of \textbf{T}, respectively,
and
\begin{equation}
\mathbf{S}=(\mathbf{e}^{(1)},\ldots ,\mathbf{e}^{(N-2)})\nonumber
\end{equation}
be the orthogonal matrix, which diagonalizes \textbf{T}, i.e.
\setcounter{equation}{16}
\begin{equation}
\mathbf{S^{-1}TS} = \mathbf{\Lambda}\,,\quad \Lambda_{\alpha\beta} =
\lambda_{\alpha}\delta_{\alpha\beta}.\label{eq14}
\end{equation}
Then we can introduce the normal coordinates
\begin{equation}
 x_{\alpha} = \sum_{i=1\atop (\neq M)}^{N-1}S_{\alpha
i}\tilde{q}_i\,,\quad p_{\alpha} = \sum_{i=1\atop (\neq
M)}^{N-1}S_{\alpha i}\tilde{p}_i\label{eq15}
\end{equation}
for $\alpha = 1, 2, ..., N-2$ such that $H_{harm}$ becomes diagonal
\begin{equation}
 H_{harm} = \frac{1}{2} \sum_{\alpha=1}^{N-2}
\left[\lambda_{\alpha}p_{\alpha}^2 + Cx_{\alpha}^2\right]\qquad
.\tag{19a}\label{eq19a}
\end{equation}
The interaction term takes the form
\begin{equation}
 H_{int} = -\sum_{\alpha=1}^{N-2}c_{\alpha}x_{\alpha}q_M\qquad
.\tag{19b}\label{eq19b}
\end{equation}
The $M$ dependent coupling constants are given by
\setcounter{equation}{19}
\begin{equation}
c_{\alpha} = C\frac{1}{m_M+m_{M+1}}\left[m_{M+1}e^{(\alpha)}_{M-1} +
m_Me^{(\alpha)}_{M+1}\right]\qquad .\label{eq17} \end{equation}
The final step is the Legendre transformation of $H$, which leads to
the Euclidean Lagrangian
\begin{equation}
L = L_d + L_1\,,\quad L_1 = L_{harm} + L_{int}\tag{21a}\label{eq21a}
\end{equation}
where
\begin{equation}
L_d = \frac{1}{2}\widetilde{m}\dot{q}_M^2 + V_0(q_M),
\tag{21b}\label{eq21b}
\end{equation}
and
\begin{equation}
L_1 = \frac{1}{2} \sum_{\alpha=1}^{N-2} m_{\alpha}
\left[\dot{x}_{\alpha}^2 + \omega_{\alpha}^2 \left(x_{\alpha} -
\frac{c_{\alpha}}{m_{\alpha} \omega_{\alpha}^2}q_M\right)^2\right]
\qquad .\tag{21c}\label{eq21c}
\end{equation}
Here we have used the equality $\tilde{q}_M=q_M$
(cf.~Eq.~(\ref{eq15b})) and
\setcounter{equation}{21}
\begin{equation}
\sum_{\alpha=1}^{N-2}\frac{c_{\alpha}^2}{m_{\alpha}\omega_{\alpha}^2}
= C\frac{m_M^2 + m_{M+1}^2}{(m_M + m_{M+1})^2}\,,\label{eq19}
\end{equation}
which follows from the completeness of the set
$\textbf{e}^{(\alpha)}$ of eigenvectors and $1/(m_{\alpha}
\omega_{\alpha}^2) = 1/C$ (see below). Eq.~(\ref{eq19}) allows us to
include the counterterm $\frac C 2 \frac {m_M^2 + m_{M+1}^2}{(m_M +
m_{M+1})^2} q_M^2$ in Eq.~(\ref{eq16b}) into $L_1$. This
counterterm, the role of which has been discussed by Caldeira and
Leggett \cite{11}, results from the canonical transformation,
Eq.~(\ref{eq15a}), (\ref{eq15b}). This transformation eliminates the
coupling between the momenta of the harmonic DOF and that of the
defect and generates coupling between the normal mode coordinates
$\{x_{\alpha}\}$ and the corresponding defect variable $q_M$
(cf.~Eq.~(\ref{eq16d})). Due to the disappearance of the
counterterm, there is no frequency renormalization for the bond
defect\cite{3}. The Lagrangian (21) is identical to that of Eq.~(4).
The masses $m_\alpha$ and frequencies $\omega_\alpha$ follow from
\begin{equation}
m_\alpha = \frac{1}{\lambda_\alpha}\,,\quad \omega_\alpha = (C
\lambda_{\alpha})^{\frac{1}{2}}\tag{23a}\label{eq23a}
\end{equation}
and the reduced defect mass $\tilde{m}$ is given by
\begin{equation}
 \widetilde{m} =
\frac{m_Mm_{M+1}}{m_M+m_{M+1}} \tag{23b}\label{eq23b}\qquad\qquad.
\end{equation}
Note that these results are exact for one dimensional systems. It
can be shown that for an \textit{arbitrary} impurity in a two- or
three-dimensional system the Langrangian (21) can be derived within
a kind of harmonic approximation \cite{12}.

\section{QUANTUM TUNNELING}
\label{tunneling}

\begin{figure}[h]
\begin{center}
\includegraphics[scale=.45]{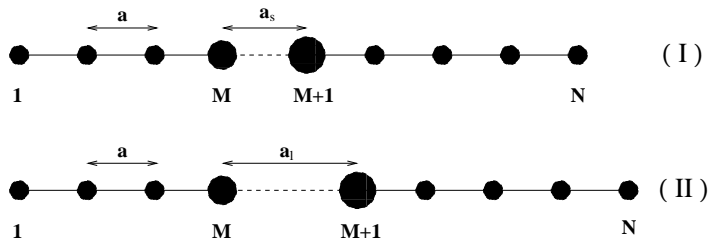}
\caption{Two degenerate classical ground states of the open chain
with $N$ particles. The masses $m_n$, $n\neq M, M+1$ are chosen to
be equal. $a$ is the equilibrium length of the harmonic bonds and
$a_s, a_l$ the two degenerate equilibrium lengths of the anharmonic
bond.}\label{fig}
\end{center}
\end{figure}

In this section we will investigate the zero temperature quantum
behavior of the anharmonic bond defect embedded in the harmonic
chain as shown in Fig. \ref{fig}.
We will assume that $V_0(q_M)$ is a symmetric double-well potential
with \textit{degenerate} minima at $q^-_M=a_s>0$ and $q_M^+ = a_\ell
>a_s$. Then the classical ground state of $V(x_1, \ldots, x_N)$
(cf.~Eq.~\ref{eq10b}) is twofold degenerate (see Fig.1). Therefore
the low-lying eigenstates form doublets. Neglecting the excited
doublets at zero temperature is justified if the bare tunneling
splitting of the ground state doublet is much less than the
frequency of the upper phonon band edge $\omega_0$ (see Eq.
(\ref{eq25a}))\cite{1}. If the total number $N$ of particles in the
chain is macroscopically large and the particle number $M$ is in the
bulk of the chain, i.e. $M = \mathcal{O}(N)$, one might have
expected a suppression of tunneling since the change from, e.g.
$q^-_M=a_s$ to $q_M^+ = a_\ell$, would require a translation of the
macroscopic mass of the left and right harmonic parts of the chain.
We will see that this naive expectation is not always correct.

On the other hand, if the defect is close to one of the free
boundaries, i.e. either $M = \mathcal{O}(1)$ or $(N-M) =
\mathcal{O}(1)$, only a finite mass $\sim M$ has to be translated.
Consequently tunneling cannot be suppressed. This qualitative
$M$-dependence should follow from that of the kernel $K(\tau)$,
which itself results from the strong $M$-sensitivity of the spectral
density $J(\omega)$. Subsection \ref{subsec_A} discusses this
phenomenon. The limit $m_M \rightarrow \infty$ (or
$m_{M+1}\rightarrow \infty$), motivated by the conclusions drawn in
Ref.~\cite{8}, will be discussed in Subsection \ref{subsec_B}.

\subsection{Location-dependent quantum tunneling}
\label{subsec_A}

As we will see in Subsection \ref{subsec_B} the form of the nonlocal
action Eq.~(\ref{eq3c}) does not depend qualitatively on the masses
$m_n$, even if one of the defect masses $m_M$ or $m_{M+1}$ becomes
infinitely large. Therefore we will choose for convenience
$m_n\equiv m$. Let us start with the situation where the bond defect
is located within the bulk, i.e. $M = \mathcal{O}(N)$, so that
\setcounter{equation}{23}
\begin{equation}
O < \lim \limits_{N \rightarrow \infty \atop M \rightarrow \infty}
\frac{M}{N} = \xi < 1.
\label{eq21}
\end{equation}
One can prove that the tunneling phenomena do not depend on $\xi$ if
it is different from 0 and 1. Therefore we choose $M=N/2$ and
without loss of generality $N$ to be even. This choice and the
assumption $m_n\equiv m$ allows us to determine the eigenfrequencies
$\omega_\alpha$ and the eigenvectors $\bold{e}^{(\alpha)},\alpha
=1,\ldots , N-2$ exactly for finite $N$. A calculation, whose
technical details are presented in Appendix \ref{A}, results in
\begin{equation}
\lambda_\alpha ^\pm = \frac 4 m \sin ^2\left(\frac
{q_\alpha^\pm}{2}\right) \rightarrow \omega _\alpha = \omega _0
\sin\left(\frac{q_{\alpha}}{2}\right)\,,\quad \omega_0=2
\sqrt{\frac{C}{m}},\tag{25a}\label{eq25a}
\end{equation}
\begin{equation*}
e^{+(\alpha)}_j = \sqrt{\frac{2}{N-1}}
\begin{cases}
\sin(q^+_{\alpha}j) \quad &,\quad 1\leq j\leq M-1 = \frac{N}{2}-1,\\
&\\
\sin(q^+_{\alpha}(j-1)) \quad &,\quad M+1 = \frac{N}{2}+ 1\leq j
\leq N-1,
\end{cases}
\end{equation*}
\begin{equation}
e^{-(\alpha)}_j = \sqrt{\frac{2}{N}}\sin(q^-_{\alpha}j)\,, \quad
1\leq j\leq N-1;\, j\neq M = \frac{N}{2}, \tag{25b}\label{eq25b}
\end{equation}
where
\begin{equation}
q_{\alpha}^{(\sigma)} =
\begin{cases}
\displaystyle\frac{\pi}{N-1}(2\alpha-1)\quad &,\quad\sigma = +,\\
&\\
\displaystyle\frac{\pi}{N}2\alpha\quad &,\quad\sigma = -
\end{cases}\tag{25c}\label{eq25c}
\end{equation}
for $\alpha = 1,\ldots, N/2-1$. It is easy to see that
$e_j^{+(\alpha)}$ are the symmetric eigenvalues with respect to $j
\rightarrow N-j$ and $e_j^{-(\alpha)}$ are the antisymmetric ones.
From this, $m _n \equiv m$ and Eq.~(\ref{eq17}) it is obvious that
the bond defect does not couple to the antisymmetric vibrational
modes, as may be expected from the symmetry of the problem. With
these results we can calculate the spectral density. Making use of
Eqs.~(\ref{eq17}), (\ref{eq23a}), (\ref{eq25a}), (\ref{eq25b}) and
(\ref{eq25c}) we get from Eq.~(\ref{eq5c}) in the thermodynamic
limit $N \rightarrow \infty$ that
\begin{equation*}
J(\omega) = \frac{1}{2}C \omega_0\int\limits_0^{\pi}dq\,
\cos^2\left(\frac{q}{2} \right) \sin\left( \frac{q}{2}\right)
\delta\Bigl(\omega - \omega_0 \sin \left(\frac{q}{2}\right)
\Bigr)
\end{equation*}
\begin{equation}
= C \sqrt{1-\left(\frac{\omega}{\omega_0}\right)^2}
\frac{\omega}{\omega_0}. \tag{26a} \label{eq26a}
\end{equation}
In the limit $\omega \ll \omega _0$ we obviously have
\begin{equation}
J(\omega) \cong C\frac{\omega}{\omega_0}\,,\tag{26b}\label{eq26b}
\end{equation}
which corresponds to the situation of Ohmic damping. The Ohmic
damping results from two facts: First, the density of states
$g(\omega)$ of the vibrational modes in a one-dimensional lattice is
constant for $\omega \ll \omega _0$ and second the squared coupling
constant involves the factor $\sin^2(q_\alpha M)$ which, for
$N\rightarrow \infty$, $M \rightarrow \infty$ with $M/N = \xi (\neq
0,1)$, oscillates faster and faster so that it can be replaced by
1/2. It is emphasized that the condition $M = \mathcal{O} (1)$, i.e.
$\xi = 0$ or 1, will change the shape of $J(\omega)$ qualitatively.

Now we can calculate the kernel $K(\tau)$. For $\omega_0T\gg 1$ and
$|\omega_0(\frac T 2 - |\tau |)|\gg 1$ the fraction in
Eq.~(\ref{eq5b}) can be well approximated by $\exp(-\omega _0|\tau |
\sin \frac q 2)$. The influence of the oscillators on tunneling of
$q_M$ is determined by the large $-\tau$ behavior, i.e. by the low
frequency modes. Therefore, substituting $J(\omega)$ from
Eq.~(\ref{eq26b}) into Eq.~(\ref{eq5b}) we find, of course, the
well-known result for Ohmic damping \cite{1,2,3}
\setcounter{equation}{26}
\begin{equation}
K(\tau) \cong \frac{1}{\pi} C \omega_0 \frac{1}{(\omega_0\tau)^2}
\,,\quad \omega_0 \tau \gg 1.\label{eq24}
\end{equation}

As a consequence, there exists a critical elastic constant
$C_{crit}$ so that the anharmonic bond can tunnel for $C <
C_{crit}$, despite macroscopic masses have to be moved (see Figure
1). For $C > C_{crit}$ symmetry is broken. If the anharmonic bond is
prepared in its ground state, e.g. $q_m^-= a_s$, it will remain
there on average.

If, however, the bond defect is located close to one of the
boundaries, so that either $M=\mathcal{O}(1)$ or $N - M =
\mathcal{O}(1)$ ($\xi =$ 0 or 1 in (\ref{eq21})), the situation
changes. Note that we perform first the thermodynamic limit
$N\rightarrow\infty$. Then $M = \mathcal{O}(1)$ means that $M$ may
equal $1, 2, ..., 10^6$ or even a larger, but still finite number.
For $m_n\equiv m$ one can easily show that
\begin{equation}
c_{\alpha} = \frac{1}{2}C\left(e^{(\alpha)}_{M-1} +
e_{M+1}^{(\alpha)}\right) =
C\mathcal{N}_{\alpha}\sin\left(q_{\alpha}M\right)\label{25}
\end{equation}
where $\mathcal{N}_\alpha$ is the normalization constant of
$\{e_n^{(\alpha)}\}$ (see Appendix \ref{A}). Replacing
$\mathcal{N}_\alpha$ by its low frequency behavior $(2/N)^{1/2}$ and
taking the limits $\omega_0T \gg 1$ and $|\omega_0(\frac T 2 - |\tau
|)| \gg 1$ yields
\begin{equation}
 K_M(\tau) \cong \frac{1}{2} C \omega_0\frac{1}{\pi}
\int\limits_0^{\pi}dq\,q \sin^2(qM) e^{-\frac{1}{2} \omega_0|\tau|q}
\label{eq26}
\end{equation}
for the kernel. In order to be consistent we also replaced
$\omega(q)=\omega_0 \sin(q/2)$ by its low frequency dispersion
$\omega(q) \cong \frac 1 2 \omega _0q$. The integrand of
Eq.~(\ref{eq26}) involves two $q$-scales
\begin{equation}
\label{eq27} q_M = \frac 1 M \quad \quad \mbox{and}\ \quad q_\tau =
\frac {1}{\omega _0|\tau|} \quad .
 \end{equation}
Equating $q_M = q_\tau$ defines the timescale
\begin{equation}\label{eq28}
\tau_M = \omega_o^{-1}M.
\end{equation}
The physical meaning of $\tau_M$ is as follows: The path integral
formalism \cite{7,13} allows one to investigate quantum tunneling by
determining the instanton solutions, i.e. the solutions of the
classical equation of motion for a double well potential in
imaginary time. The width $\tau_{kink} $ of a single instanton is
\begin{equation*}
\tau_{kink} = \left(\frac{m}{V''_0(q_M^\pm)}\right)^{1/2}\quad .
\end{equation*}
If we assume that $V''_0(q_M^\pm)\approx C$, the elastic constant of
the harmonic bonds, then $ \tau_{kink}\approx \omega_0^{-1} $ so
that
\begin{equation}\label{eq29}
\tau_M \approx M\tau_{kink}\quad .
\end{equation}
The $\tau$-dependence of $K_M(\tau)$ is sensitive to whether
$|\tau|/\tau_{kink} \gg M$ or vice versa. Let us start with the long
time limit
\begin{itemize}
\item[(i)] $|\tau | / \tau_M \gg
1$.
\vspace{.5cm}

Then it follows from Eqs.~(\ref{eq27}) and (\ref{eq28}) that
$q_\tau\ll q_M$. The major contribution to the integral in
Eq.~(\ref{eq26}) comes from $q < q_\tau \ll q_M$. Therefore, we are
allowed to replace $\sin^2(qM) = \sin^2(q/q_M)$ by $(q/q_M)^2 =
(qM)^2$, which leads to the spectral density $J(\omega) \sim
\omega^3$ at low frequencies corresponding to {\it superohmic}
damping. It implies that $K_M(\tau) \sim \tau^{-4}$. The precise
result is
\begin{equation}
K_M(\tau) \cong \frac{48}{\pi}C\omega_0 \frac{1}{(\omega_0\tau)^4}
M^2\,, \quad |\tau| \gg\tau_M\quad .\label{eq30}
\end{equation}
\begin{figure}[h]
\includegraphics[scale=.275]{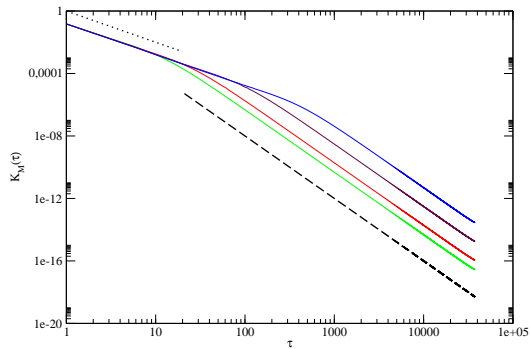}
\caption{$\tau$-dependence of $K_M(\tau)$ for $M = 5, 10, 40$ and
$160$ (from bottom to top) on a log-log representation. The dotted
and dashed line, corresponds to $\tau^{-2}$ and $\tau^{-4}$,
respectively. The crossover at $\tau \approx \tau_M$ from the
$\tau^{-2}$ behavior to that of $\tau^{-4}$ can clearly be seen. The
parameters have been chosen as follows: $T=10^5, C = m = 1
\Rightarrow \omega_0 = 2$.}
\end{figure}
\item[(ii)] A different situation takes place for $|\tau|/\tau_M \ll 1$.

Then $q_\tau \gg q_M$ and the integral in Eq.~(\ref{eq26}) must be
decomposed in two contributions $\int\limits _0^{q_M}dq \cdots +
\int \limits ^\pi _{q_M} dq \cdots$. The first integral yields a
constant in the leading order in $\omega_0\tau/\tau_M =
\omega_0\tau/M$. For the second one we are allowed to replace
$\sin^2(q M) = \sin^2(q/q_M)$ by $1/2$ since the main contribution
comes from $q \approx q_{\tau} \gg q_M$ so that the function
$\sin^2(qM) = \sin^2(q/q_M)$ is oscillating fast between zero and
one, whereas $\exp\left(- \frac{1}{2}\omega_0|\tau|q\right) =
\exp\left(- \frac{1}{2}q/q_{\tau}\right)$ varies slowly. Accordingly
$\tau \ll \tau_M$ corresponds to an 'effective' spectral density
$J(\omega) \approx \omega$, i.e. to Ohmic damping. It is
straightforward to estimate both integrals. As the final result we
obtain
\begin{equation}
K_M(\tau) \cong \frac{1}{\pi}C\omega_0 \Biggl[\frac{1}{8M^2}
\biggl(1+\mathcal{O} \left(\frac{\omega_0|\tau|}{M}
\right)\biggr)+\frac{1}{(\omega_0\tau)^2}\Biggr]\label{eq31}
\end{equation}

for $\tau _{kink} \ll \tau \ll \tau_M  \approx M \tau _{kink}$.
Taking the limit $M\rightarrow \infty$ in Eq.~(\ref{eq31}) restores
the result (\ref{eq24}) for the kernel $K(\tau)$.
\end{itemize}

For $M = \mathcal{O}(1)$ or $N-M = \mathcal{O}(1)$, and
$N\rightarrow\infty$ we obtain for $K_M(\tau)$ a crossover at
$\tau_M$ from the power law ~ $\tau^{-2}$ for $\tau \ll \tau_M$ to
$\tau^{-4} $ for $\tau \gg \tau_M$. Figure 2 illustrates this
behavior for $K_M(\tau)$, calculated numerically. On the log-log
plot of Figure 2 the crossover between both power laws can easily be
observed.

This crossover is related to the $M$-dependence of the spectral
density because the defect-phonon coupling constants $c_{\alpha}$
are $M$-dependent. If the 'observation time' $T$ (which is the time
$T$ in $G_E$ from Eq. (\ref{eq2})) is smaller than $\tau_M$ then
$(\tau-\tau')$ will be smaller than $\tau_M$, as well. Consequently
the kernel $K_M(\tau-\tau')$ entering the nonlocal action (Eq.
(\ref{eq3c})) decays as $1/(\tau-\tau')^2$. If, however, $T$ is
larger than $\tau_M$ then it is possible that $(\tau-\tau')$ becomes
larger than $\tau_M$, as well. For those values of $(\tau-\tau')$
the kernel $K_M$ decays as $1/(\tau-\tau')^4$. This discussion
reveals that the choice of the 'observation time' $T$ allows to fix
the 'large'-$\tau$ behavior of $K_M(\tau)$, where, of course, $\tau
\leq T$. If $M$ is far away from the chain end the crossover time
$\tau_M$ is correspondingly large. Increasing $M$ even more $\tau_M$
increases, too. Nevertheless, the 'observation time' dependence
still exists. It disappears for $M\rightarrow\infty$, only. We
remind the reader that the limit $N\rightarrow\infty$ has to be
taken first.

Now making use of the analogy \cite{1,3,4} between the calculation
of the action of a multi-instanton configuration interacting via
$K_M(\tau - \tau ')$ and a one-dimensional Ising model with the
coupling constants $J_{nm}\sim K_M(n-m) \sim |n-m|^{-k}$ we can
conclude the following: If the 'observation time' $T$ is smaller
than $\tau _M$ then we have $K_M(\tau) \sim \tau^{-2}$, i.e.
\textit{Ohmic} damping. In that case the bond defect may tunnel for
$C < C_{crit}(T)$, whereas symmetry becomes broken for
$C>C_{crit}(T)$. Note, that this is not a sharp transition at
$C_{crit}(T)$ since finite $T$ corresponds to a finite Ising chain
which does not exhibit a sharp phase transition. What really happens
when increasing the coupling constant $C$ is an increase of the
correlation 'length' $\xi(C)$. As soon as $\xi(C)$ equals the 'size'
$T$ of the Ising-chain a 'long range' order occurs. However, if $T$
is much larger than $\tau_M\approx M\tau _{kink}$ we have
$K_M(\tau)\sim \tau ^{-4}$ and tunneling is never suppressed
\cite{1,3}.

One might be puzzled by these conclusions since the transition for
Ohmic damping to decoherence for $M = \mathcal{O}(N), N\rightarrow
\infty$ occurs for large $T$, or to be more precise it becomes a
sharp transition for $T = \infty$, only. As we already stressed
above, the transition for $T < \tau_M$ is {\em not} sharp. The
relevant phonons contributing to $K_M(\tau)$ have wavenumbers
$q\approx q_{\tau} \sim q_T = \frac{1}{T}$. This makes the
'effective' spectral density Ohmic. Mapping the situation for $M =
\mathcal{O}(1)$ again onto the Ising chain of length $L$ results in
the coupling constants $J_{nm}$ decaying like $J_0|n - m|^{-2}$ for
$|n-m| < M$ and like $J_0|n-m|^{-4}$ for $|n- m| > M$. It is clear
that there is {\em no} sharp phase transition for finite $L$. But if
$L < M$, the coupling constants decay as $J_0|n-m|^{-2}$. For a
fixed temperature $T$ (not to be confused with the 'observation
time' $T$) there will be no magnetic order ($\hat{=}$ coherent
tunneling) for $J_0 < J_{0,crit}(T) (\hat{=}\,C<C_{crit}(T))$. For
$J_0 > J_{0,crit}(T)(\hat{=}\,C>C_{crit}(T))$ a crossover to the
'long range order' ($\hat{=}$ decoherent tunneling) takes place.
Accordingly the quantum tunneling phenomenon is richer for $M =
\mathcal{O}(1)$ than for $M=\mathcal{O}(N)$.

Actually we may think also in the real time $t$ terms that as long
as only the phonons with relatively high frequencies $\omega \approx
1/t$ and hence short wave length such that $1/q < M$ participate in
the interaction with the anharmonic defect, the latter does not
'feel' that the chain is finite and behaves as in the Ohmic case. In
the course of time the lower frequency phonons with higher wave
length 'reach' the end of the chain and a crossover to a superohmic
behavior takes place.

\subsection{Dependence on the masses of defect}
\label{subsec_B}

In this subsection we will assume that
\begin{equation}
m_n =
\begin{cases}
m\quad &,\quad n\neq M, M+1,\\
M_1\quad &,\quad n = M,\\
M_2\quad &,\quad n = M+1.
\end{cases}\label{eq32}
\end{equation}
For $M = \mathcal{O}(N)$ we may choose without loss of generality $M
= N/2$ and $N$ being even. It is easy to prove that $e_j^{+
(\alpha)}$ from Eq.~(\ref{eq25b}) remain eigenvectors of \textbf{T}
with $q^+_\alpha$, $\alpha = 1,2,\ldots, N/2-1$ given by
Eq.~(\ref{eq25c}). The remaining $(N/2-1)$ eigenvectors are of the
form
\begin{equation}
e^{-(\alpha)}_j = \mathcal{N}_{\alpha}^-
 \begin{cases}
\sin(q^-_{\alpha}j) \quad &,\quad 1\leq j\leq\frac{N}{2}-1,\\
& \\
\sin(q^-_{\alpha}(N-j)) \quad &,\quad \frac{N}{2}+1\leq j \leq
N-1\quad.
 \end{cases}\tag{36a}\label{eq36a}
\end{equation}
$q_\alpha ^-$ is a solution of transcendental equation. Let us
introduce the quantities $\beta _i = m/M_i$, $i = 1,2$ and $\delta =
\beta _1 \beta _2/(\beta _1 + \beta _2)$. The limit $M_1 \rightarrow
\infty $ (or $M_2 \rightarrow \infty)$ implies $\delta \rightarrow
0$. Since the results in \cite{8} motivate us to study, e.g. $M_2
\rightarrow \infty$, we find
\begin{equation}
q_{\alpha}^- = q_{\alpha}^+ + \frac{1}{N-1}
\frac{2\delta}{\tan(\frac{q^+_{\sigma}} {2})} +
\mathcal{O}(\delta^2)\tag{36b}\label{eq36b}
\end{equation}
in the limit $\delta \rightarrow 0$. Substituting
Eqs.~(\ref{eq36a}), (\ref{eq36b}) into Eq.~(\ref{eq17}) yields
\begin{equation*}
c_\alpha^-\cong c_\alpha ^+\quad .
\end{equation*}
in the leading order in $\delta$. As a result $J(\omega)$ and
therefore $K(\tau)$, too, are doubled as compared to the case of
$M_i = m$, i.e. we get
\setcounter{equation}{36}
\begin{equation}\label{eq34}
K(\tau)\cong \frac 2 \pi C \omega_0 \frac {1}{(\omega _0\tau)^2}
\quad .
\end{equation}
for $M_2\to \infty$.

The only essential result of changing $M_2$ from $m$ to infinity is
that the critical elastic constant increases by a factor of two.

\section{Summary and Conclusions}
\label{summary}

For a translationally invariant chain with one anharmonic bond and
otherwise harmonic nearest neighbor interactions, we have shown
exactly how the anharmonic degree of freedom can be separated from
the harmonic ones in their normal mode representation. As a result,
we have obtained Lagrangian (21), which is of the form of Lagrangian
(4). Note, that this result can also be obtained for a
three-dimensional system within the Born-Oppenheimer approximation
starting with an arbitrary translationally invariant potential
$V(\vec{x}_1, \ldots,\vec{x}_N)$ for a $N$-particle system
\cite{12}. Since the Caldeira-Leggett type nonlocal action
\cite{1,2,3} is based on the form (4) (or (21)) of the Lagrangian it
is not the translation invariance and therefore not the conservation
of momentum, which can lead to a different type of nonlocal action.
The discrepancy between our results and those of Ref.~\cite{8} may
have the following origin. Since the harmonic part of Lagrangian
(\ref{eq6}) is diagonal in $x_\alpha$, the harmonic variables
$x_\alpha$ are already normal mode coordinates. In that case a
translation of the full system only changes the Goldstone mode
amplitude, let us say $x_0$, but leaves all the other normal mode
coordinates unchanged, i.e. $x_\alpha \rightarrow x_\alpha$, $\alpha
\neq 0$ for any translation. If $x_1$ and $x_2$ in Eq.~(\ref{eq6})
are real space coordinates then the coupling term $(x_\alpha -
x_2)^2$ is not translationally invariant for $\alpha \neq 0$.

Our model has allowed us to calculate explicitly, e.g. for $m_n  =
m$ and $M = N/2$ the coupling constants $c_\alpha$, the
eigenfrequencies $\omega_\alpha$ and the spectral density
$J(\omega)$. The $\omega$-dependence of $J$ is determined by the
density of states $g(\omega)$ and the coupling constants $c_\alpha$.
Although $g(\omega)$ for $\omega \rightarrow 0$ is independent of
the number $M$, the frequency dependence of $c_\alpha$ exhibits a
sensitivity to the location $M$, which makes the affect of the
harmonic bath on quantum tunneling $M$ sensitive. As a consequence,
the damping is Ohmic if the bond defect is within the bulk of the
chain and superohmic if it is close to the boundaries. For the
former case there is a transition from a delocalized state (due to
tunneling) to a localized one if the elastic constant exceeds a
critical value $C_{crit}$, whereas tunneling is never suppressed in
the latter case, provided the 'observation time' $T$ is large enough
compared to $\tau_M$ which is roughly $M$ times the instanton kink
width. For $T < \tau_M$ (since the thermodynamic limit
$N\rightarrow\infty$ had already been performed, $M$ must be finite,
but can be arbitrary large) the dissipation is effectively Ohmic
leading to a similar behavior when the bond defect is within the
bulk.

If $M = \mathcal{O}(N)$ (e.g. $M = N/2$) and if one of the masses of
the bond defect tends to infinity, no significant changes occur
except for doubling of the critical constant $C_{crit}$. This is
obvious since, e.g. $M_2 \to \infty$, makes the part of the chain to
the right of the defect inactive, i.e. the phonons to the right do
not act as a reservoir for the bond defect. Accordingly, only half
of the harmonic chain is generating dissipation, which results in
doubling of $C_{crit}$.

Although we are not aware of a concrete experimental system, these
results could be relevant for linear macromolecules, which may be
described by the model Hamiltonian Eq. (10). If many of such
molecules with a single defect are produced. the position of which
can be controlled experimentally, one might observe, e.g. the
location-dependent tunneling by spectroscopic methods.

\bigskip

\underline{Acknowledgement} This work has been completed when two of
us (V.F. and R.S.) were a member of the Advanced Study Group 2007 at
the MPIPKS Dresden. V.F. and R.S. gratefully acknowledge the MPIPKS
for its hospitality and financial support. V.F. acknowledges a
support of Israeli Science Foundation, Grant N 944/05. One of us
(R.S.) thanks Eugene Chudnovsky for helpful discussions.
\appendix
\section{Use of COM and relative coordinates of the total
chain: Diagonalization of the matrix {\bf T}}
\label{A}

The separation of the harmonic and anharmonic DOF by using the
center of mass and relative coordinates of the {\em total} chain has
been described in Section \ref{Model}. The transformation to normal
coordinates requires the diagonalization of the matrix {\bf T}. Here
the most important steps of the diagonalization procedure are
outlined.

Making use of Eqs. (12)-(15) one obtains the matrix elements of the
symmetric matrix {\bf T} in the form
\begin{equation}
T_{ii} =
\begin{cases}
\displaystyle \frac{m_i + m_{i+1}}{m_im_{i+1}} &,\quad i=1, ...,
M-2, M+2, ...,
N-1,\\
& \\
\displaystyle \frac{m_{M-1}+(m_M + m_{M+1})}{m_{M-1}(m_M+m_{M+1})}
&,\quad i=M-1,
\\
& \\
\displaystyle \frac{(m_M+m_{M+1})+m_{M+2}}{(m_M+m_{M+1})m_{M+2}}
&,\quad i=M+1,\end{cases}\tag{A1a} \label{eqA1a}
\end{equation}
\begin{equation}
T_{ii+1} =
\begin{cases}
- \displaystyle \frac{1}{m_{i+1}} &,\quad i=1, ..., M-2, M+1, ...,
N-2,
\\
&\\ 0 &,\quad i = M-1\end{cases}\tag{A1b}\label{eqA1b}\\
\end{equation}
and
\begin{equation}
T_{ii+2} =
\begin{cases}
0 &,\quad i=1, ..., M-3, M+1, ..., N-3,\\
&\\ -\displaystyle\frac{1}{m_M+m_{M+1}} &,\quad i=M-1\quad
.\tag{A1c}\label{eqA1c}\end{cases}
\end{equation}
The diagonalization of {\bf T} can not be done analytically for
arbitrary masses $m_i$. Therefore we take the simplest case of equal
masses, $m_i \equiv m$. Then it is straightforward to prove that the
eigenvalue equation
$$
\sum_{j=1\atop (j\neq M)}^{N-1}T_{ij}e^{(\alpha)}_j =
\lambda_{\alpha}e^{(\alpha)}_i
$$
is solved by
\begin{equation}
e_i^{(\alpha)} = \mathcal{N}_{\alpha}
\begin{cases}\sin(q_{\alpha}i),
& 1\leq i\leq M-1, \\
& \\
b_{\alpha}\sin(q_{\alpha}(N-i)), & M+1\leq i \leq N-1,
\end{cases}\tag{A2}\label{eqA2}
\end{equation}
\begin{equation}
\lambda_{\alpha} =
\frac{2}{m}(1-\cos(q_{\alpha}))\tag{A3}\label{eqA3}
\end{equation}
where $\mathcal{N}_{\alpha}$ is the normalization constant and
$b_{\alpha}$ is a coefficient depending on the location $M$ of the
bond defect. The wave numbers $q_{\alpha}$ are solutions of the
transcendental equation
\begin{equation}
\cot(qN) = \cot(qM)+\frac{\cot(\frac{q}{2})}{2\sin^2(qM)}\quad
.\tag{A4}\label{eqA4}
\end{equation}
Since the l.h.s. of Eq. (\ref{eqA4}) diverges at $q = \frac{\pi}{N}
\cdot \alpha,\,\alpha=0,1,...$ it is easy to see that its solutions
are of the form
\begin{equation}
q_{\alpha} = \frac{\pi}{N}\alpha + \epsilon_{\alpha},\quad \alpha=0,
1, ..., N-3\tag{A5}\label{eqA5}
\end{equation}
with $0\leq \epsilon_{\alpha} < \frac{\pi}{N}$. There are $(N-2)$
solutions corresponding the $(N-2)$ harmonic DOF. The remaining two
DOF are the COM and the bond defect coordinate $X_c$ and $q_M$,
respectively. In the thermodynamic limit $N\rightarrow\infty$, the
discrete set of $q_{\alpha}$ wave vectors becomes a continuous
variable $q$ within $[0,\pi]$ with a constant density which together
with Eq. (\ref{eqA3}) implies a constant low energy density of
states.

The normalization constant $\mathcal{N}_{\alpha}$ and the
coefficient $b_{\alpha}$ are functions of $q_{\alpha}, M$ and $N$.
Their explicit expressions are not given here.

\section{Use of COM and relative coordinates of the bond defect}
\label{B}

In this appendix we will describe the separation of harmonic and
anharmonic DOF using an approach alternative to that used in Section
\ref{Model}. It has the advantage that it can be straightforwardly
applied to higher dimensional systems. The starting point is the
introduction of COM and relative coordinate $X_d$ and $q_M$,
respectively, of the {\it bond defect} (see Eqs. (\ref{eq11a}) and
(\ref{eq11b})). The corresponding canonical momenta
\begin{equation}
P_d = p_M + p_{M+1}\tag{B1a}\label{eqB1a},
\end{equation}
\begin{equation}
\pi_M = \frac{m_M}{m_M+m_{M+1}} p_{M+1} -
\frac{m_{M+1}}{m_M+m_{M+1}}p_M. \tag{B1b}\label{eqB1b}
\end{equation}
Substituting $X_d,\ q_M$ from Eq. (11) and $P_d,\ \pi_M$  from (B1)
into Eq. (10) yields
\begin{equation}
H = H_d + H_{harm} + H_{int}\tag{B2a}\label{eqB2a}
\end{equation}
where
\begin{equation}
H_d = \frac{1}{2\mu_M}\pi_M^2 + V_0(q_M) +
\frac{C}{2}\frac{m_M^2+m_{M+1}^2}{(m_M+m_{M+1})^2}q_M^2 \tag{B2b}
\label{eqB2b}
\end{equation}
\begin{equation*}
H_{harm} = \sum_{n=1\atop (n\neq M, M+1)}^{N}\frac{1}{2m_n}p_n^2 +
\frac{1}{2(m_M+m_{M+1})}P_d^2+\frac{C}{2}\sum_{n=1\atop (n\neq M,
M\pm 1)}^{N-1}(x_{n+1}-x_n - a_n)^2
\end{equation*}
\begin{equation}
+ \frac{C}{2}(X_d-x_{M-1} - a_{M-1})^2 +
\frac{C}{2}(x_{M+2}-X_d-a_{M+1})^2\tag{B2c}\label{eqB2c}
\end{equation}
\begin{equation}
H_{int} = - C\left[ \frac{m_{M+1}}{m_M+m_{M+1}}(X_d-x_{M-1}-a_{M-1})
+ \frac{m_M}{m_M+m_{M+1}}(x_{M+2}-X_d-a_{M+1})\right]q_M\quad
.\tag{B2d}\label{eqB2d}
\end{equation}
Here $\mu_M = m_M m_{M+1}/(m_M+m_{M+1})$ is the reduced mass of bond
defect. Note that $H_d$ from Eq. (\ref{eqB2b}) is identical to $H_d$
from Eq. (\ref{eq16b}) after replacing $(\pi_M, q_M)$ by
$(\tilde{p}_M,\tilde{q}_M)$. The transformation of $H_{harm}$ (Eq.
(\ref{eqB2c})) to the normal coordinates may be more conveniently
carried out using the notations
\begin{equation}
x'_n = \begin{cases}
x_n &,\quad n=1, ..., M-1\\
X_d &,\quad n= M\\
x_{n+1} &,\quad n=M+1, ..., N-1
\end{cases}
\tag{B3a}\label{eqB3a}
\end{equation}
\begin{equation}
p'_n = \begin{cases}
p_n &,\quad n=1, ..., M-1\\
P_d &,\quad n= M\\
p_{n+1} &,\quad n=M+1, ..., N-1
\end{cases}\tag{B3b}\label{eqB3b}
\end{equation}
and
\begin{equation}
m'_n = \begin{cases}
m_n &,\quad n=1, ..., M-1\\
m_M+m_{M+1} &,\quad n= M\\
m_{n+1} &,\quad n=M+1, ..., N-1
\end{cases}\quad .\tag{B3c}\label{eqB3c}
\end{equation}
Next we expand the potential part $V_{harm}(x'_1, ..., x'_{N-1})$ of
$H_{harm}$ around its equilibrium configuration,
\begin{equation}
x'_n = x'^{(eq)}_n + u'_n,\tag{B4}\label{eqB4}
\end{equation}
up to the second order terms in $u'_n$. Note that this is not an
approximation since $V_{harm}$ is a harmonic potential. This leads
to
\begin{equation}
H_{harm} = \sum_{n =1}^{N-1}\frac{1}{2m'_n}p'^2_n
+\frac{C}{2}\sum_{n=1}^{N-2}(u'_{n+1}-u'_n)^2\quad
.\tag{B5}\label{eqB5}
\end{equation}
Introducing the mass-weighted coordinates
\begin{equation}
\tilde{u}'_n = \sqrt{m'_n}u'_n,\tag{B6a}\label{B6a}
\end{equation}
\begin{equation}
\tilde{p}'_n = \frac{1}{\sqrt{m'_n}}p'_n,\tag{B6b}\label{eqB6b}
\end{equation}
Eq. (\ref{eqB5}) yields
\begin{equation}
H_{harm} = \frac{1}{2} \sum_{n=1}^{N-1}\tilde{p}'^2_n +\frac{1}{2}
\sum_{n,m = 1}^{N-1} \widetilde{V}'_{nm}\tilde{u}'_n
\tilde{u}'_m\tag{B7a}\label{eqB7a}
\end{equation}
where the only nonzero matrix elements of the symmetric matrix
$\mathbf{\widetilde{V}}'$ are
\begin{equation}
\widetilde{V}'_{nn} = \frac{C}{m'_n}\begin{cases}
1 &,\quad n=1, N-1\\
2 &,\quad n=2, ..., N-2
\end{cases}
\tag{B7b}\label{eqB7b}
\end{equation}
and
\begin{equation}
\widetilde{V}'_{nn+1} = - \frac{C}{\sqrt{m'_nm'_{n+1}}}.
\tag{B7c}\label{eqB7c}
\end{equation}

Let $\tilde{e}^{(\alpha)}_n$ and $\tilde{\lambda}_{\alpha}, \,
\alpha=0, 1, ..., N-2$ be respectively the eigenvectors and
eigenvalues of $\mathbf{\widetilde{V}'}$. The canonical
transformation
\begin{equation}
\tilde{x}_{\alpha} =
\sum_{n=1}^{N-1}\tilde{u}'_n\tilde{e}^{(\alpha)}_n,\quad
\tilde{p}_{\alpha} =
\sum_{n=1}^{N-1}\tilde{p}'_n\tilde{e}^{(\alpha)}_n\tag{B8}
\label{eqB8}
\end{equation}
leads to the normal mode representation
\begin{equation}
H_{harm} = \frac{1}{2} \sum_{\alpha=0}^{N-2}
\left[\tilde{p}^2_{\alpha} + \tilde{\lambda}_{\alpha}
\tilde{x}^2_{\alpha}\right].\tag{B9} \label{eqB9}
\end{equation}
Since $H_{harm}$ in Eq. (\ref{eqB5}) is still translation invariant
there is a zero frequency mode which we choose for $\alpha = 0$.
With $\lambda_0 = 0$ we get
\begin{equation}
H_{harm} = \frac{1}{2}\tilde{p}^2_0 +
\frac{1}{2}\sum_{\alpha=1}^{N-2} \left[\tilde{p}^2_{\alpha} +
\tilde{\lambda}_{\alpha}\tilde{x}^2_{\alpha}\right]\quad
.\tag{B10}\label{eqB10}
\end{equation}
The first term in the r.h.s. of Eq. (\ref{eqB10}) is the kinetic
energy of the COM of {\em total} chain. The second term corresponds
to $H_{harm}$ from Eq. (\ref{eq19a}). Using Eqs. (B3), (\ref{eqB4}),
(B6) and (\ref{eqB8}) brings the interaction part (Eq.
(\ref{eqB2d})) into the form
\begin{equation}
H_{int} = -\sum_{\alpha}\tilde{c}_{\alpha}\tilde{x}_{\alpha}
q_M\tag{B11}\label{eqB11}
\end{equation}
with
\begin{equation}
\tilde{c}_{\alpha} = C \frac{1}{m_M+m_{M+1}} \left[m_{M+1}
\left(\frac{1}{\sqrt{m'_M}} \tilde{e}^{(\alpha)}_M -
\frac{1}{\sqrt{m'_{M-1}}}\tilde{e}^{(\alpha)}_{M-1}\right) + m_M
\left(\frac{1}{\sqrt{m'_{M+1}}} \tilde{e}^{(\alpha)}_{M+1} -
\frac{1}{\sqrt{m'_M}} \tilde{e}^{(\alpha)}_M \right)\right]\tag{B12}
\label{eqB12}
\end{equation}
Again, the analytical diagonalization of $\mathbf{\widetilde{V}'}$
cannot be performed for arbitrary masses. Accordingly, we choose
$m_n \equiv m$ as in Appendix \ref{A}. Then Eqs. (\ref{eqB7b}) and
(\ref{eqB7c}) result in
\begin{equation}
m'_n = m\begin{cases}
1 &,\quad n\neq M,\\
2 &,\quad n=M,
\end{cases}\tag{B13}\label{eqB13}
\end{equation}
\begin{equation}
\widetilde{V}'_{nn} = \frac{C}{m}\begin{cases}
1 &,\quad n=1, M, N-1,\\
2 &,\quad n\neq 1, M, N-1,
\end{cases}
\tag{B14a}\label{eqB14a}
\end{equation}
\begin{equation}
\widetilde{V}'_{nn+1} = -\frac{C}{m}\begin{cases}
1 &,\quad n\neq M-1, M,\\
1/\sqrt{2} &,\quad n=M-1, M
\end{cases}\quad .
\tag{B14b}\label{eqB14b}
\end{equation}
All the other matrix elements vanish. Again it is straightforward to
prove that the eigenvalue equation
$\sum_{m=1}^{N-1}\widetilde{V}'_{nm}\tilde{e}^{(\alpha)}_m =
\tilde{\lambda}_{\alpha}\tilde{e}^{(\alpha)}_n$ for $n \neq M-1, M,
M+1$ is solved by
\begin{equation}
\tilde{e}^{(\alpha)}_n =
\widetilde{\mathcal{N}}_{\alpha}\begin{cases}
\cos\left(\tilde{q}_{\alpha}\left(n - \frac{1}{2}\right)\right)
&,\quad n=1, ..., M-2,
\\
\tilde{b}_{\alpha}\cos\left(\tilde{q}_{\alpha}
\left(N-n-\frac{1}{2}\right)\right) &,\quad n=M+2, ..., N-1,
\end{cases}
\tag{B15}\label{eqB15}
\end{equation}
\begin{equation}
\tilde{\lambda}_{\alpha} =
\frac{2C}{m}(1-\cos(\tilde{q}_{\alpha}),\tag{B16}\label{eqB16}
\end{equation}
with $\widetilde{ \mathcal{N}}_{\alpha}$ being the normalization
constant and $\tilde{b}_{\alpha}$ a $M$-dependent coefficient. The
remaining equations for $\tilde{e}^{(\alpha)}_n$ with $n=M-1, M$ and
$M+1$ yield a nontrivial solution if a corresponding determinant
vanishes. This condition leads to the transcendental equation
\begin{equation}
2(-1+2\cos(q)) = \frac{\cos\left(q\left(M -
\frac{3}{2}\right)\right)}{\cos\left(q\left(M -
\frac{1}{2}\right)\right)} + \frac{\cos\left(q\left(N - M -
\frac{3}{2}\right)\right)}{\cos\left(q\left(N - M -
\frac{1}{2}\right)\right)}\tag{B17}\label{eqB17}
\end{equation}
for the wave numbers $\tilde{q}_{\alpha}$. Although Eq.
(\ref{eqB17}) looks quite different from the transcendental equation
(\ref{eqA4}) it can be shown by use of identities for trigonometric
functions that (\ref{eqB17}) and (\ref{eqA4}) are equivalent, i.e.
the set of solutions $\{\tilde{q}_{\alpha}\}$ of Eq. (\ref{eqB17})
and $\{q_{\alpha}\}$ of Eq. (\ref{eqA4}) are identical. We have
already stressed that $H_d$ from Eq. (\ref{eqB2b}) and that from Eq.
(\ref{eq16b}) are identical, as well. Straightforward but tedious
calculations show that the complete Lagrangian corresponding to the
classical Hamiltonian Eq. (B2) is identical to the Lagrangian (21).
Particularly, it can be proven that $\tilde{c}_{\alpha}$ from Eq.
(\ref{eqB12}) is identical to $c_{\alpha}$ from Eq. (\ref{eq17}).

\section{Separating the harmonic part into left and right parts}
\label{C}

In this appendix we will show that separation of the harmonic and
anharmonic DOF can be done by taking the left and right harmonic
parts separately. Similarly to the approach used in Section
\ref{Model} our first step is to separate the COM of the {\em total}
chain from the relative coordinates. This leads to the Hamiltonian
from Eq. (\ref{eq11}). Neglecting the kinetic energy of COM Eq.
(\ref{eq11}) can be rewritten as
\begin{equation}
H = H_d + H^L_{harm} + H^R_{harm} + H_{int}\tag{C1a}\label{eqC1a}
\end{equation}
where
\begin{equation}
H_d = \frac{1}{2\mu_M}\pi_M^2 + V_0(q_M),\tag{C1b}\label{eqC1b}
\end{equation}
\begin{equation}
H^L_{harm} = \frac{1}{2}\sum_{i,j=1}^{M-1}T_{ij}^L\pi_i\pi_j +
\frac{C}{2}\sum_{i=1}^{M-1}q_i^2,\tag{C1c}\label{eqC1c}
\end{equation}
\begin{equation}
H^R_{harm} = \frac{1}{2}\sum_{i,j=M+1}^{N-1}T_{ij}^R\pi_i\pi_j +
\frac{C}{2}\sum_{i=M+1}^{N-1}q_i^2,\tag{C1d}\label{eqC1d}
\end{equation}
\begin{equation}
H_{int} = - \left(\frac{1}{m_M}\pi_{M-1} +
\frac{1}{m_{M+1}}\pi_{M+1}\right) \pi_M \tag{C1e}\label{eqC1e}
\end{equation}
and the nonzero matrix elements are
\begin{equation}
T_{ii}^{(\sigma)} = \frac{m_i + m_{i+1}}{m_im_{i+1}},\quad
T_{ii+1}^{(\sigma)} = -\frac{1}{m_{i+1}} =
T_{i+1,i}^{(\sigma)},\quad \sigma=L, R\tag{C1f}\label{eqC1f}
\end{equation}
with $i=1, ..., M-1$ for $\sigma=L$ and $i=M+1, ..., N-1$ for
$\sigma=R$. Let $e^{L(\nu)}_i(e^{R(\mu)}_i)$ and
$\lambda_{\nu}^L(\lambda_{\mu}^R)$ be the eigenvectors and
eigenvalues of $\mathbf{T}^L(\mathbf{T}^R)$.

Then we use the notations
\begin{equation}
x_{\nu}^L = \sum_{i=1}^{M-1}q_i e^{L(\nu)}_i,\quad x_{\mu}^R =
\sum_{i=M+1}^{N-1}q_i e^{R(\mu)}_i\tag{C2a}\label{eqC2a}
\end{equation}
\begin{equation}
p_{\nu}^L = \sum_{i=1}^{M-1}\pi_i e^{L(\nu)}_i,\quad p_{\mu}^R =
\sum_{i=M+1}^{N-1}\pi_i e^{R(\mu)}_i\tag{C2b}\label{eqC2b}
\end{equation}
in order to get
\begin{equation}
H^L_{harm} = \frac{1}{2}\sum_{\nu=1}^{M-1}\left[\lambda_{\nu}^L
(p_{\nu}^L)^2 + C(x_{\nu}^L)^2\right],\tag{C3a}\label{eqC3a}
\end{equation}
\begin{equation}
H^R_{harm} = \frac{1}{2}\sum_{\mu=M+1}^{N-1}\left[\lambda_{\mu}^R
(p_{\mu}^R)^2 + C(x_{\mu}^R)^2\right]\tag{C3b}\label{eqC3b}
\end{equation}
for the harmonic part and
\begin{equation}
H_{int} = -\left(\sum_{\nu=1}^{M-1}c_{\nu}^Lp_{\nu}^L
+\sum_{\mu=M+1}^{N-1}c_{\mu}^Rp_{\mu}^R
\right)\pi_M\tag{C4a}\label{eqC4a}
\end{equation}
for the interaction with
\begin{equation}
c_{\nu}^L = \frac{1}{m_M}e^{L(\nu)}_{M-1},\quad c_{\mu}^R =
\frac{1}{m_{M+1}}e^{R(\mu)}_{M+1}\tag{C4b}\label{eqC4b}
\end{equation}
for the coupling constants.

This type of approach describes the chain as a bond defect coupled
to {\em two} baths of harmonic oscillators, the left and right part
of the chain. For the path integral formalism we need the
Lagrangian. From (\ref{eqC1a}), (\ref{eqC1b}), (C3) and
(\ref{eqC4a}) we can determine the velocities $\dot{q}_M,
\dot{x}_{\nu}^L$ and $\dot{x}_{\mu}^R$ as function of the momenta.
Solving for the momenta as a function of the velocities is
straightforward but tedious. We report the final result
\begin{equation}
\pi_M = \kappa\left[\dot{q}_M +
\sum_{\nu=1}^{M-1}\frac{c_{\nu}^L}{\lambda_{\nu}^L}\dot{x}_{\nu}^L +
\sum_{\mu=1}^{N-M-1} \frac{c_{\mu}^R}{\lambda_{\mu}^R}
\dot{x}_{\mu}^R\right]\tag{C5a}\label{eqC5a}
\end{equation}
\begin{equation}
p_{\nu}^L = \frac{1}{\lambda_{\nu}^L}\dot{x}_{\nu}^L + \kappa
\frac{c_{\nu}^L}{\lambda_{\nu}^L}
\left[\sum_{\nu'=1}^{M-1}\frac{c_{\nu'}^L}{\lambda_{\nu'}^L}
\dot{x}_{\nu'}^L + \sum_{\mu'=1}^{N-M-1}
\frac{c_{\mu'}^R}{\lambda_{\mu'}^R}\dot{x}_{\mu'}^R +
\dot{q}_M\right]\tag{C5b}\label{eqC5b}
\end{equation}
\begin{equation*}
p_{\mu}^R = \frac{1}{\lambda_{\mu}^R}\dot{x}_{\mu}^R +
\kappa\frac{c_{\mu}^R}{\lambda_{\mu}^R}
\left[\sum_{\nu'=1}^{M-1}\frac{c_{\nu'}^L}{\lambda_{\nu'}^L}
\dot{x}_{\nu'}^L +
\sum_{\mu'=1}^{N-M-1}\frac{c_{\mu'}^R}{\lambda_{\mu'}^R}\dot{x}_{\mu'}^R
+ \dot{q}_M\right]\tag{C5c}\label{eqC5c}
\end{equation*}
with
\begin{equation}
\kappa = \mu_M \left[1 - \mu_M
\left(\sum_{\nu=1}^{M-1}\frac{(c_{\nu}^L)^2}{ \lambda_{\nu}^L} +
\sum_{\mu=1}^{N-M-1}\frac{(c_{\mu}^R)^2}{
\lambda_{\mu}^R}\right)\right]^{-1}.\tag{C6}\label{eqC6}
\end{equation}
Making use of (C5) for the calculation of the Legendre transform of
$H$ from Eq. (C1) leads to the Euclidean Lagrangian
\begin{equation}
L = L_d + L_{harm} + L_{int}\tag{C7a}\label{eqC7a}
\end{equation}
where
\begin{equation}
L_d = \frac{\kappa}{2}\dot{q}_M^2 + V_0(q_M)\tag{C7b} \label{eqC7b}
\end{equation}
in which
\begin{equation*}
L_{harm} = \frac{1}{2} \sum_{\nu=1}^{M-1}
\left[\frac{1}{\lambda_{\nu}^L} (\dot{x}_{\nu}^L)^2 +
C(x_{\nu}^L)^2\right] + \frac{1}{2} \sum_{\mu=1}^{N-M-1} \left[
\frac{1}{\lambda_{\mu}^R}(\dot{x}_{\mu}^R)^2 +
C(x_{\mu}^R)^2\right],
\end{equation*}
\begin{equation}
+ \frac{\kappa}{2} \left[\sum_{\nu=1}^{M-1}
\frac{c_{\nu}^L}{\lambda_{\nu}^L}\dot{x}_{\nu}^L +
\sum_{\mu=1}^{N-M-1} \frac{c_{\mu}^R}{\lambda_{\mu}^R}
\dot{x}_{\mu}^R\right]^2, \tag{C7c}\label{eqC7c}
\end{equation}
\begin{equation}
L_{int} = -\kappa\, \dot{q}_M \left[ \sum_{\nu=1}^{M-1}
\frac{c_{\nu}^L}{\lambda_{\nu}^L} \dot{x}_{\nu}^L +
\sum_{\mu=1}^{N-M-1} \frac{c_{\mu}^R}{\lambda_{\mu}^R}
\dot{x}_{\mu}^R\right].\tag{C7d}\label{eqC7d}
\end{equation}
This form of $L$ differs completely from that of Eq. (21).
Particularly, $L_{int}$ from Eq. (\ref{eqC7d}) is a coupling of the
velocities and not of the bond defect coordinate $q_M$ with the
normal mode coordinates $x_{\alpha}$ as for $L_{int}$ from Eqs.
(\ref{eq21a}), (\ref{eq21c}). In addition, the harmonic part Eq.
(\ref{eqC7c}) is not 'diagonal', i.e. due to the third term on the
r.h.s. of Eq. (\ref{eqC7c}) there is an intra- and an inter-coupling
between the phonons (normal modes) of the left and right harmonic
part of the chain.

In order to eliminate the harmonic degrees of freedom in the path
integral representation of the propagator one has to 'diagonalize'
$L_{harm}$ from Eq. (\ref{eqC7c}). This can be done by a point
transformation $x_{\nu}^L(\{x_{\alpha}\},q_M)$ and $ x_{\mu}^R(
\{x_{\alpha}\}, q_M) $. This transformation follows directly from
Eqs. (\ref{eq15b}), (18) and (C2):
\begin{equation}
x_{\nu}^L(\{x_{\alpha}\},q_M) = \sum_{\alpha=1}^{N-2}
\left(\sum_{i=1}^{M-1} e_i^{(\alpha)} e^{L(\nu)}_i\right) x_{\alpha}
- \frac{m_{M+1}}{m_M+m_{M+1}} e^{L(\nu)}_{M-1}
q_M\tag{C8a}\label{eqC8a}
\end{equation}
\begin{equation}
x_{\mu}^R(\{x_{\alpha}\},q_M) = \sum_{\alpha=1}^{N-2}\left(
\sum_{i=M+1}^{N-1}e_i^{(\alpha)}e^{R(\mu)}_i\right)x_{\alpha} -
\frac{m_M}{m_M+m_{M+1}} e^{R(\nu)}_{M+1} q_M\tag{C8b}\label{eqC8b}
\end{equation}
Taking the time derivative of Eq. (C8) yields the transformation of
the velocities. Substituting this and the transformation (C8) into
Eq. (C7) 'diagonalizes' $L_{harm}$ and replaces the velocity
coupling by a coupling of $q_M$ and $\{x_{\alpha}\}$. After a
lengthy calculation one arrives at the Lagrangian from Eq. (21),
which, of course, is not a surprise.

\end{document}